**Software paper for submission to the Journal of Open Research Software**

To complete this template, please replace the blue text with your own. The paper has three main sections: (1) Overview; (2) Availability; (3) Reuse potential.

Please submit the completed paper to: editor.jors@ubiquitypress.com

---

## (1) Overview

### Title
A Web-based modeling tool for the SEMAT Essence theory of Software Engineering

### Paper Authors
1. Graziotin, Daniel, daniel.graziotin@unibz.it
2. Abrahamsson, Pekka, pekka.abrahamsson@unibz.it

### Paper Author Roles and Affiliations
1. Project development, Free University of Bozen-Bolzano
2. Project manager, Free University of Bozen-Bolzano

### Abstract
As opposed to more mature subjects, software engineering lacks general theories to establish its foundations as a discipline. The Essence Theory of software engineering (Essence) has been proposed by the Software Engineering Methods and Theory (SEMAT) initiative. Essence goal is to develop a theoretically sound basis for software engineering practice and its wide adoption. Essence is yet far from reaching academic and industry adoption. Reasons include a struggle to foresee its utilization potential and the lack of tools implementing it. SEMAT Accelerator (SematAcc) is a Web-positioning tool for a software engineering endeavor, which implements the SEMAT's Essence kernel. SematAcc allows using Essence, thus helping to understand it. The tool enables teaching, adopting, and researching Essence in controlled experiments and case studies.

### Introduction
Well-established academic disciplines emerged from practices without focusing on the underlying theory [1]. Along with time and research activities, however, general and advanced scientific theories have been developed as they are essential for the advancement of scientific fields [2]. Software engineering is notably a young discipline [3]. As such, it is not yet overly concerned with a core, general theory; thus, the risk is to be limited to trial-and-error practices [4]. A lack of theoretical foundations inhibits the growth of a cumulative research tradition, which is essential no matter if core knowledge has to be built or if scientific revolutions occur [2]. Software engineering suffers from gaps in the knowledge for understanding software development processes and their impact on each other [5].

The SEMAT initiative was born in order to "support a process to refound software engineering based on a solid theory, proven principles and best practices. [...][*What will enable this is a*] kernel of widely-agreed elements [...] supported by industry, academia, researchers and users" [6]. The outcome of the SEMAT initiative is the *Essence Theory of Software Engineering* [7,8].

Strictly speaking, Essence documents the "things to work with" in software engineering [8], the relationship they have with each other, and actions Concerning the "things".

Essence is claimed to provide a common basis for defining software development practices, by using widely agreed elements that are present in every software engineering endeavor [8]. These elements are called *Alphas*. As of today, the core Alphas of Essence kernel are Opportunity, Stakeholders, Requirements, Software System, Work, Team, and Way-of-working. Alphas change in their *States*, thus enabling a representation of the progress and health of the endeavor. For example, the Requirements can be Conceived, Bounded, Coherent, Acceptable, Addressed, or Fulfilled. Finally, Essence Alphas are organized in three areas of *Concern*. Each Concern focuses on a single aspect of software engineering. They are called Customer, Solution, and Endeavor.

As opposed to other attempts to create a general theory of software engineering – e.g. the Software Engineering Body of Knowledge [9] – Essence aims to generalize software engineering by identifying its universal elements and actions, and to develop a universal language to describe them. The theory was submitted to the Object Management Group (OMG) and is currently undergoing the necessary steps to become an OMG standard [8]. A simple introduction to Essence is available in the material of a special lecture held at the Free University of Bozen-Bolzano by one of the authors of this paper [10].

As of today, there is far from consensus acceptance of Essence as a model. For example, notable practitioners and researchers of Agile methodologies expressed negative comments on the SEMAT initiative (e.g., [11–13]). Reasons include a difficulty to see Essence utilization potential and the lack of tools implementing it. It is challenging to understand Essence before even evaluating it. As software engineering heavily relies on empirical research [14], there is the need to produce the empirical data when applying Essence, in order to understand its theoretical and practical implications and to evaluate it scientifically.

This paper describes the SEMAT Accelerator (SematAcc), which is a Web-based modeling system for software engineering processes as depicted by the SEMAT's Essence kernel. SematAcc has been developed as a way to produce such needed empirical data for evaluating Essence. It helps in using and understanding the Essence theory under a practical viewpoint. SematAcc lets users model a software development process with Essence elements. A software engineering endeavor (called *project* in SematAcc) has an associated Essence kernel. The kernel has been implemented using the OMG submission of Essence [8] as a reference. A software engineering endeavor is represented using graphs of the Concerns' and Alphas' completions.

SematAcc is developed in JavaScript both on the client side and on the server side, on top of the recently born Meteor [15]. Meteor is an open-source platform to build JavaScript-based Web applications. By providing a nearly identical API for the development of the server and the client, Meteor target is to deliver almost real-time data transfer through latency compensation techniques and reactive programming.

SematAcc can be employed in empirical experiments and case studies on Essence theory because it registers the events triggered by its usage. The events can be easily downloaded as a CSV string. Thus, they are directly employable in statistical software.

The rest of this paper is organized as follows. The next section - Implementation and Architecture - provides a brief introduction to Meteor, an illustrated guide on how to use SematAcc main functionality, and more technical details on the architecture of the system and the source-code organization. It is followed by a section regarding the Quality Control of the system, where the tests and the techniques in order to ensure a high quality of the delivered software are provided. In the Availability section, details on SematAcc system requirements and dependencies are given, as well as the information on how to obtain the software. The last section - Reuse Potential - provides an overview on how to configure and run SematAcc on a local computer and on a remote server, and suggestions on how to employ the software for research purposes.

**Implementation and architecture**

SematAcc was implemented using the recent Meteor project [15]. Meteor is an open-source platform and framework to build JavaScript-based Web applications. It is built upon the server-side JavaScript enabling technology Node.js [16] and the NoSQL document-oriented MongoDB database [17].

As a programming framework, Meteor permits developers to write both client- and server-side applications using almost the same JavaScript APIs. It is its aim to achieve nearly real-time performance by providing latency compensation for data transfers between the server and clients (and vice versa). Data is the central point of Meteor. It is organized in data structure named collections. Meteor collections are convenient wrappers around MongoDB documents. They provide the basic create, read, update, and delete operations needed for the functionality of a Web application.

Meteor has been inspired by the reactive programming paradigm [18]: nearly all of its layers, from database access to graphical user interface (GUI), provide an event-driven interface whose operations run whenever the underlying data-dependencies change. For example, a template of the GUI of a Meteor application is responsible to render a list of books. Suppose a user is seeing the list of books.

Meanwhile, an editor inserts a new book in the list from the backend. The change in the data is detected by Meteor components, which automatically trigger a re-render of the GUI element interested by this operation. The user simply sees a new element added to the list, without a Web page refresh.

As an infrastructure, Meteor covers development, testing, and production phases to run the implemented source-code. For development, Meteor provides a local server, built on top of Node.js [16], to run the developed Website and perform tests. For testing and production, Meteor provides a freely available infrastructure of servers to be employed for testing and production purposes. More information on Meteor can be obtained on the official Website [15], in the introductory screencast [19], and on the documentation Website [20].

Before illustrating some details regarding SematAcc architecture, a brief introduction on using SematAcc is provided. Upon login, the user manages the projects (Figure 1). The Hint Box on the left side of the screen contains suggestions to help the user manage the projects. The central box lists the available projects. After the user creates a project, putting the mouse over it will activate the available commands. A click on the arrow icon will accelerate a project with SematAcc (that is, it will start the modelling activity with Essence kernel). The other two commands are for editing a project and for deleting it.

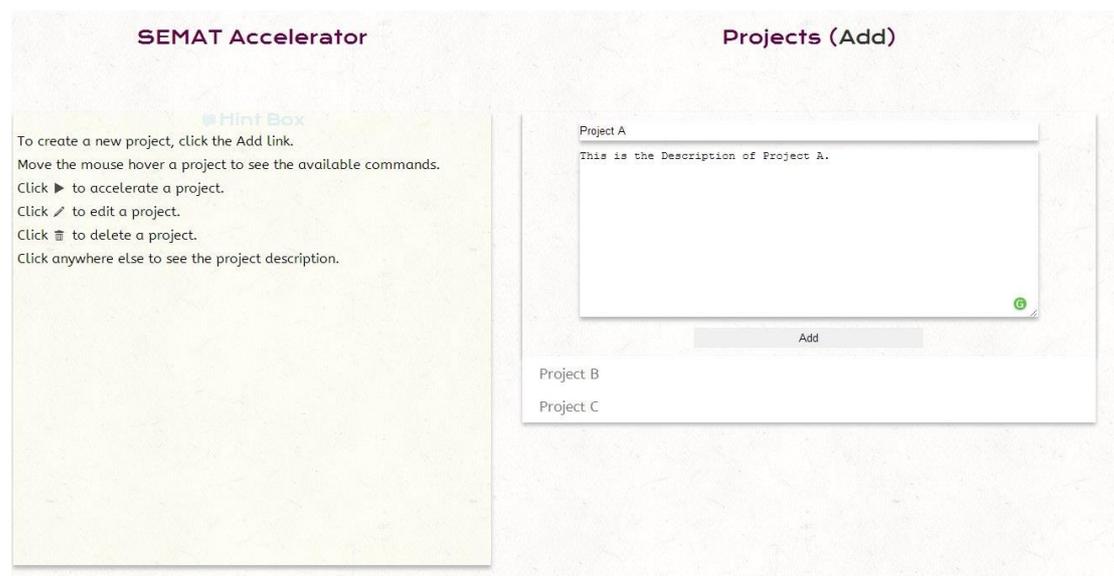

*Figure 1 Project Management in SematAcc*

When the user decides to accelerate a project with SematAcc, the Essence kernel is loaded with the chosen project (Figure 2). This is the main functionality of SematAcc. The user interface is divided in three parts. The first part, on the left, is the already mentioned Hint Box, which reports information on an Alpha or a State, depending on the position of the mouse. The second part, on the center, is Essence kernel. It shows the Alphas. If the user clicks on an Alpha, the Kernel expands and shows the corresponding States. The third part, on the right, is the project status. It displays a rose graph and a horizontal bar chart. The rose graph represents the project status in terms of Alphas' completions, whereas the horizontal bar chart provides an overview of the Concerns' completion.

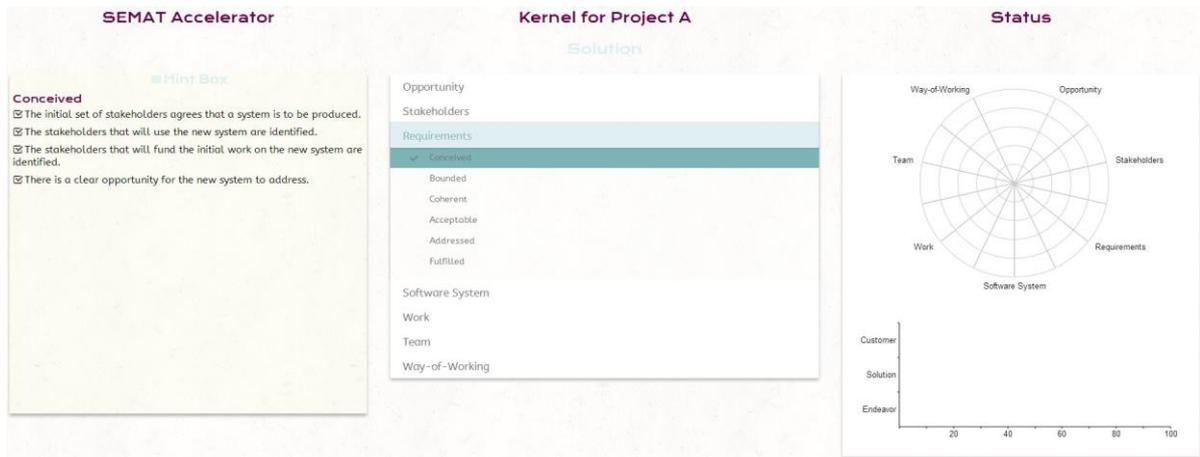

*Figure 2 SematAcc Main Window: the Hint Box, the Essence kernel, and the graphs.*

When a user clicks on a State, thus expressing a change of State of an Alpha, the graphs are immediately updated. For example, in Figure 3, the user choses that Requirements Alpha are now in the Conceived State. The rose graphs and the horizontal bar chart on the right side of Figure 3 represent the change in the State of an Alpha, and the completion of the project is shown.

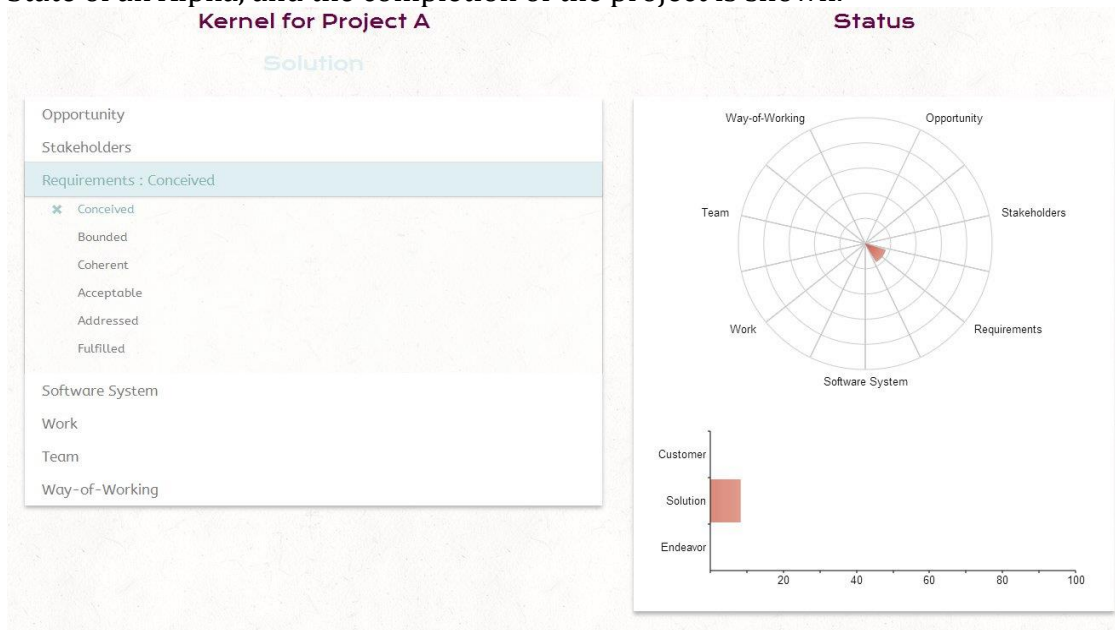

*Figure 3 Change of State in SematAcc.*

The models of SematAcc derive from an analysis of the Essence theory. They are represented in the class diagram of Figure 4.

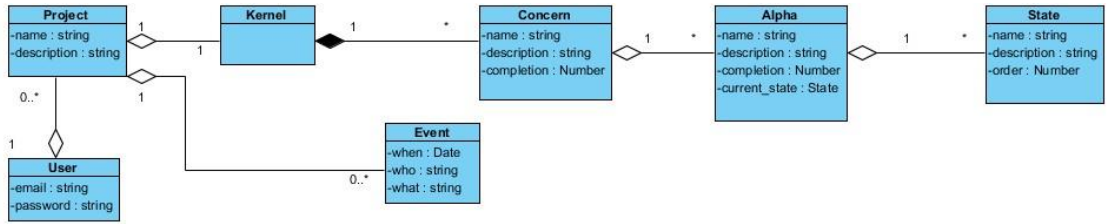

Figure 4  SematAcc Models

A user of the system possesses one-to-many projects. Each project has an associated Essence kernel. In SematAcc, the kernel is a graphical conceptualization and does not have a corresponding database collection. Although a kernel may have one-to-many Concerns, the current OMG standard proposal suggests three of them: "Customer", "Solution", and "Endeavor".  The three Concerns contain the corresponding Alphas. Alphas have several associated States but may be in zero or one-and-only-one State.

Each Concern, Alpha, and State has a corresponding name and description taken from Essence OMG submission. In order to speed-up the Essence learning process, the descriptions of kernel elements appear in SematAcc as soon as the user hovers with the mouse over them.

Concerns and Alphas also possess a completion, represented as a JavaScript Number. This numbers represent the percentage of completion of such kernel elements. For Alphas completion, the order of the current Alpha State is employed. Then, the completion of the Alphas determines the progress of the project in terms of Concerns completion.

Finally, each project possesses *events*, which are employed by SematAcc to log and generate data for research purposes.

A high-level view of the most important functionality of SematAcc is represented in Figure 5.

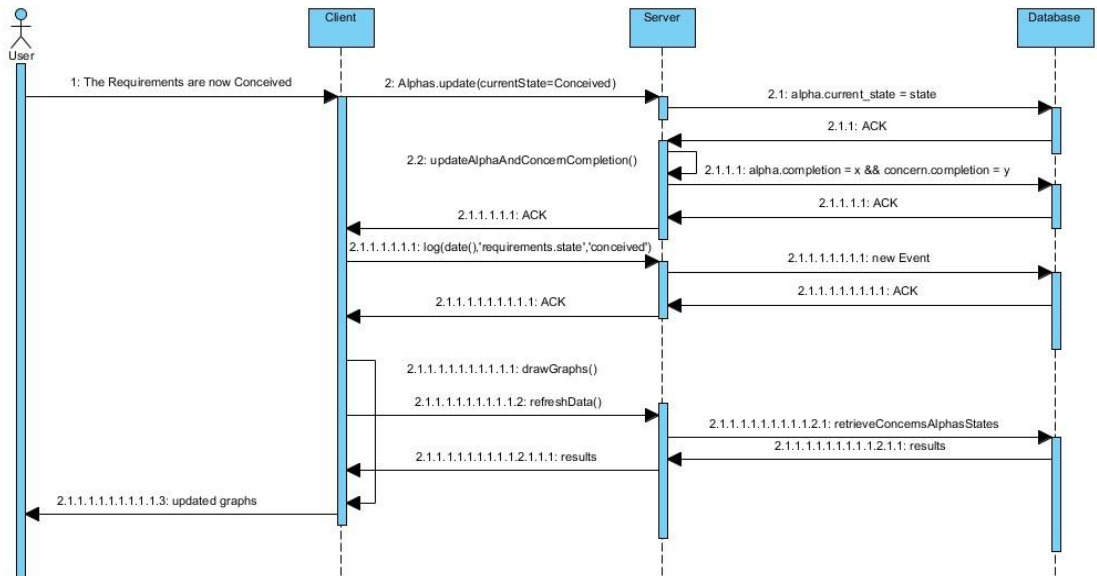

Figure 5  High-level view of a change in Alpha State

In the example provided in Figure 3, the user decides the "Requirements" are now "conceived". A mouse click on the "conceived" State in the user interface triggers an update to the server's Alpha collection. Upon a successful update in the database, the completion ratios for the Concerns and the Alphas are recalculated in order to keep the data consistent. Then, the client component of SematAcc is notified of the successful change of State operation. A relevant event is generated and stored. Finally, the graphs are updated to represent the new data. At this state, SematAcc GUI shows what it is represented in Figure 3.

The project source-code is organized by separation of concerns. The project's root directory contains three main parts, according to Meteor conventions. In the *client* folder, there is the code executed by the client (the Web browser). In the *server* folder lives the source-code, which is purely executed on the server side. The *tests* folder hosts the automated tests written with the Laika framework [21]. The *public* folder stores all the binary files publicly accessible – e.g., pictures. All the source-code files are commented. For more information regarding the project structure, we advise to read the README file in SematAcc source-code. The file also contains instructions on how to run SematAcc on a development machine and how to have an own private instance of SematAcc on Meteor's free servers.

**Quality control**

Different levels of testing have been performed. Eight volunteers have been involved to perform user-acceptance tests of the software. Three of them are experts of Essence. The remaining five are experts in the field of software engineering but needed an introduction to Essence theory. Their involvement persisted up to the first public release of SematAcc. Their feedback influenced the development of Essence, especially of its GUI.

Before a public release, SematAcc is manually tested with the latest versions of the leading browsers (i.e., Mozilla Firefox, Google Chrome, Microsoft Internet Explorer, and Apple Safari).

An official automated testing method does not exist for Meteor yet [22]. However, community-based tools are under active development. SematAcc employs the recently born Laika [21], a feature rich testing framework for Meteor, which simulates the interaction between the server and clients. The tests are run before each commit to the version control system. In order to run the tests, the Laika framework has to be installed on the system. More information on how to run the test cases can be found in the *TESTING.md* file in SematAcc root directory.

Lastly, each SematAcc function has been carefully debugged. Formal inspections of the database content revealed consistent data values after the software trials. The JavaScript code adheres to standard coding conventions and passes all JS Hint tests [23].

**(2) Availability**

*Operating system*

Server-side: SematAcc works wherever Meteor works.
Official support [24] is offered for:
- Mac OS X 10.6 and above
- GNU/Linux x86 and x86_64

Microsoft Windows is non-officially supported and not immediately updated [25].
Client-side: any operating system, which runs a recent Web-browser.

### *Programming language*
HTML, CSS, and JavaScript over Meteor framework v0.6.4.

### *Additional system requirements*
A minimum resource requirement is the ability to run the Meteor framework, i.e. *Node.js* [16] and *MongoDB* [17].
In a GNU/Linux development machine, the memory map of the processes, gathered with the *pmap* command, was of 295.02Mb.
SematAcc source code occupies about 1Mb of disk space.

### *Dependencies*
Meteor v0.6.4 [15].

### *List of contributors*
Graziotin, Daniel, daniel.graziotin@unibz.it
Abrahamsson, Pekka, pekka.abrahamssson@unibz.it

### *Software location:*
#### *Archive*
**Name:** figshare
**Persistent identifier:** http://dx.doi.org/10.6084/m9.figshare.664127
**Licence:** 3-clause BSD license
**Publisher:** Daniel Graziotin under a 3-clause BSD license
**Date published:** 02/04/2013
#### **Code repository**
**Name:** GitHub
**Identifier:** https://github.com/s4fs/sematacc
**Licence:** 3-clause BSD license
**Date published:** 09/10/2013

### *Language*
English.

**(3) Reuse potential**

Before describing the reuse potential of SematAcc, the instructions on how to obtain and run SematAcc are provided.

For demonstration purposes, an example instance of SematAcc is available (http://sematacc.meteor.com). Users need to register to the system in order to employ it. However, a demo project is ready to be used (http://sematacc.meteor.com/demo) without requiring registration. The demo project is not owned by any user and is shared between all users of the system. Changes made to the demo project are immediately pushed to any client that is visualizing the demo project. It must be noted that the example instance is for testing the software, and it is not for production environment. While attempts are made to preserve the data at each new version deployment, this cannot be guaranteed.

Preferably, SematAcc should run on a local machine or on Meteor's freely available servers. In the following paragraphs, we describe both cases.

SematAcc needs Meteor as unique dependency. Any other dependency has been included in SematAcc source-code. Meteor is officially available for GNU/Linux and Mac OS X, while official support for Microsoft Windows is under development [24]. The installation of Meteor is straightforward for the supported platform as it is reduced to a single line to be input in a terminal. Please see the Quick start section in Meteor Documentation [20]. For Microsoft Windows, non-official support is provided. The instructions and an installer for Microsoft Windows are provided in Meteor for Windows [25].

After the installation of Meteor, the command *meteor* has to be issued in a terminal, within the root folder of SematAcc. A Meteor instance will run and serve the project. SematAcc will be accessible from a Web browser at *http://localhost:3000*. In order to test-deploy SematAcc on Meteor's free servers, the command *meteor deploy –P chosenname.meteor.com* has to be issued from the root folder of SematAcc. After the choosing of a password, SematAcc will automatically be deployed and will be available at *http://chosenname.meteor.com*, given that the chosen name is available. These instructions are valid for all operating systems.

Support requests for SematAcc are accepted and welcomed through the issue tracker of the project (https://github.com/s4fs/sematacc/issues). Bug reports and feature requests should be entered in the issue tracker. Other enquiries can be made via e-mail to the authors of this project.

Regarding the reuse potential, SematAcc serves for multiple purposes in research activities on Essence theory. The tool has been developed specifically for producing empirical data on Essence and it is currently suitable only for this purpose. SematAcc may be employed as a tool for setting up case studies in the adoption of the Essence theory. How practitioners can learn the Essence theory can be studied. SematAcc can also be employed in empirical experiments and case studies when a software development progression is measured in terms of State transitions and Concern visualization.

Each change in an Alpha State is recorded and can be exported as a CSV string. The example of Figure 3 would generate an event of the form *<"2013-04-03T14:01:27.007Z","Requirements.State", "Conceived">.* The events are directly exportable as a CSV string from the graphical user interface. Other events can be easily recorded by modifying the source-code of SematAcc, as a single line of code is required to record an Event. See the README file of the project source-code for more information.

**Acknowledgements**

We kindly acknowledge Dr. Ivar Jacobson, who provided his constructive feedback for SematAcc while discussing the Essence theory with us. We would like to thank all the anonymous users who tried the system and provided user-acceptance data for it. We are grateful for the insightful feedback received by two anonymous reviewers, which helped us to improve the paper significantly.

**Funding Statement**

This project is financially supported by the PhD grant of Free University of Bozen Bolzano.